\documentclass[twocolumn,prb,showpacs,floatfix]{revtex4}  
\usepackage{graphicx}
\usepackage{verbatim}
\usepackage{amssymb}

\usepackage{latexsym}

\def\beeq{\begin{equation}}
\def\eneq{\end{equation}}
\def\beeqa{\begin{eqnarray}}
\def\eneqa{\end{eqnarray}}

\begin{document}
\DeclareGraphicsExtensions{.ps,.ps,.eps}

 \title{Self-vacancies in Gallium Arsenide: an {\it ab initio} calculation}
 
 \author{Fedwa El-Mellouhi} \email {f.el.mellouhi@umontreal.ca}
\author{Normand Mousseau } \email {Normand.Mousseau@umontreal.ca}
 \affiliation{D\'epartement de physique and Regroupement qu\'eb\'ecois
   sur les mat\'eriaux de pointe, Universit\'e de Montr\'eal, C.P.
   6128, succ. Centre-ville, Montr\'eal (Qu\'ebec) H3C 3J7, Canada}

\date{\today}

\begin{abstract}
  We report here a reexamination of the static properties of
  vacancies in GaAs by means of first-principles density-functional
  calculations using localized basis sets. Our calculated formation
  energies yields results that are in good agreement with recent
  experimental and {\it ab-initio} calculation and provide a complete
  description of the relaxation geometry and energetic for various
  charge state of vacancies from both sublattices.  Gallium vacancies
  are stable in the 0, $-$, $-2$, $-3$ charge state, but $V_{Ga}^{-3}$
  remains the dominant charge state for intrinsic and n-type GaAs,
  confirming results from positron annihilation.
  Interestingly, Arsenic vacancies show two successive negative-U
  transitions making only $+1$ , $-1$ and $-3$ charge states stable, while
  the intermediate defects are metastable.  The second transition
  ($-/-3$) brings a resonant bond relaxation for $V_{As}^{-3}$
  similar to the one identified for silicon and GaAs divacancies. 
\end{abstract}


\maketitle
\section{Introduction}
\label{sec:intro}

Native point defects are involved in virtually every process during
which an atom incorporated in the lattice of a semiconductor migrates
toward another lattice site. This diffusion mediated by point defects
is responsible for a number of important effects, for instance, those
encountered during fabrication of microelectronic devices. It is not
surprising therefore that point defects in semiconductors have been
extensively studied using {\it ab initio} techniques. Some work still
remains to be done, however, especially  in the case of alloyed
semiconductors. 

If Ga vacancies in GaAs are relatively well understood --- because
most dopants used in technology (Si donor, Zn, Be, and Mg acceptors)
occupy the Ga sites--- much less is known about the As vacancy. The
introduction of carbon as a prospective As-site acceptor causes a
growing interest in this defect, however. There is therefore an
abundant literature on calculations of formation energies of point
defects in GaAs which is nicely reviewed and summarized in a paper of
Deepak {\it et al.}~\cite{Dee03}. Since formation energies are difficult to
measure, calculations, are the primary method for obtaining
these values.  However because of the assumptions and approximations
taken into account, the reported values in the literature during the
last two decades differ greatly from paper to paper.

Most calculations, were done for small supercells. These calculations
take advantage of error cancellations to obtain energy differences
that are more precise than the total energies themselves. Here we
repeat these calculations for all possible charge states for both Ga
and As vacancies using the strict convergence criteria and a large
simulation cell.

The outline of the paper is as follows.  We first describe the method
for defect calculation in section~\ref{sec:simu}, then we turn to the
convergence tests made to set up the methodology. We focus on the
effect of the $k$-point sampling, system-size and completeness of the
basis on the formation energy and structural properties of the
defects. Section~\ref{sec:effects} deals with these effects on the
most important charge states like $V_{As}^{-1}$ and $V_{Ga}^{-3}$. In
section~\ref{sec:results} we present and discuss the most converged
results using localized basis sets compared to previous results
obtained from theory and experiment.

\section{Simulation details and methodology}
\label{sec:simu}
\subsection{Total energy calculations}
\label{sec:total}

The total energies for this work are evaluated using
SIESTA~\cite{SAN97, Soler02}, a self-consistent density functional
method (DFT) within local-density approximation (LDA).  Core electron
are represented by the standard norm-conserving Troullier-Martins
 pseudo-potentials~\cite{Trou91} factorized in the Kleiman-Baylander
form~\cite{Klei82} and the one-particle problem is solved using linear
combination of pseudo-atomic orbitals (PAO) basis set of finite range.
These orbitals are strictly localized and represent well the local
electronic densities; few of them are therefore needed, decreasing
considerably the computational costs by comparison with standard
plane-waves calculations.  The main drawback of this approach,
however, is the lack of a systematic procedure to ensure a rapid
variational convergence with respect to the number of basis orbitals
and to the range and shape of each orbital. Consequently, while
extending plane-wave basis sets is trivial, some efforts are needed to
prepare unbiased pseudo-atomic basis sets (see, for example,
Ref~\onlinecite{Jun01, Ang02}).

In this work, we use the following sequence to test the convergence of
the basis set. Starting with the simplest scheme, a single $\zeta$
(SZ) basis, a second group of valence orbitals is added for
flexibility, forming the double-$\zeta$ (DZ) basis. For completeness,
we also add polarization orbitals to both valence sets, generating
SZP and DZP bases.

Finally, it is possible to optimize the localization radius in order
to increase accuracy for a given basis set. While the computational
efficiency is slightly reduced, as the optimized orbitals have
generally a longer tail, it is often a good alternative to increasing
the size of the basis set (for details see Ref.~\onlinecite{SAN97,
  Soler02, Jun01}).
We also test the accuracy of these basis sets optimized with respect
to the amount of overlap between atomic orbitals around the defect
using the optimizing procedure of Anglada et al.~\cite{Ang02} at 0.0
GPa for the SZ basis set. We find that the efficiency of these
orbitals with (SZP-O) and with (SZ-O) is comparable to those of DZP
and DZ, respectively.

\begin{table}
\caption{
  Comparison between  converged  basic parameters for bulk
  GaAs. a, B and $E_{total}$ represent the lattice
  constant (\AA), the bulk modulus (GPa), and the total energy per GaAs
  pair  (eV/pair), respectively. $E_g$ denotes the energy gap (eV) and
  $\Delta H$ is the heat of formation of GaAs calculated 
  using Eq.~\ref{eq:heat} (eV). See the text for description of the
  basis set label. Results are compared to a recent plane-wave
  calculation (PW)~\cite{Zol03}  and to experimental values from
  Ref.\onlinecite{Lev96} at 0K, unless other references are cited.
} 
\label{tab:basics}
\begin{ruledtabular}
\begin{tabular} {llllllll}
&SZ &SZ-O &SZP-O &DZ &DZP  &PW &Expt.\\
\\
\hline
{\bf a} &5.68 &5.66 &5.6 &5.64 &5.6 &5.55 &5.65\\
{\bf B} &59.3 &68.9 &78.8 &67.7 &70.4 & &75.3\\
{\bf $E_{total}$} &235.2 &235.5 &235.9 &235.7 &236.0\\
{\bf $E_g$ }&0.61 &0.78 &0.98 &0.66 &0.82 &$1.08^a$ &1.54\\
&&&&&& $0.7^b$ \\
{\bf $\Delta H$ } &0.66 & 0.99 &0.78 &0.81 &0.72 &$1.0^a$ &$0.73^e$\\
&&&&&&$0.67^c$ \\
&&&&&& $0.83^d$\\
\end{tabular}
\end{ruledtabular}

\begin{tabular}{lll}
$^a$ Reference~\onlinecite{Zol03}& 
$^b$ Reference~\onlinecite{Laa92}&
$^c$ Reference~\onlinecite{Jan03} \\
$^d$ Reference~\onlinecite{Che94}&
$^e$ Reference~\onlinecite{Wea92}
\end{tabular}
\end{table}

Table~\ref{tab:basics} reports the values of a number of structural
and thermodynamical quantities for bulk GaAs as computed using these
various bases with a $k$-point sampling density of 0.03\AA$^{-1}$,
corresponding to that for a 216-atom unit cell with a $2\times 2\times
2$ $k$-point sampling. For SZ, the lattice constant at zero pressure
is found to be 5.68 \AA, overestimating the experimental value by
only 0.03 \AA. The density increases with the size of the basis set and the
lattice constant for DZP is found to be too small by 0.05 \AA\ with
respect to experiment.  The relatively contracted structure obtained
with DZP is characteristic of LDA; plane-wave calculations also using
LDA give 5.55 \AA.~\cite{Zol03} Increasing the basis set leads to a
significant improvement on the calculated value of the bulk modulus as
it goes from 59.85 GPa for SZ to 70.4 GPa for DZP, close to the
experimental value of 75.3 GPa.

The LDA band gap is found to be 0.61 eV, 0.66 eV and 0.82 eV for SZ,
DZ and DZP basis sets respectively, underestimating, as usual with
with this approximation, the experimental gap of 1.54 eV. The DZP
band gap lies well within the range of energy gaps obtained from PW
(0.7-1.0 eV),~\cite{Zol03,Laa92} however, and can be considered
converged.

No such systematic problem is found for the total energy and the heat
of formation. In particular, the heat of formation obtained with DZ
and DZP is very close to the experimental value~\cite{Wea92}, showing
a better agreement than previous plane-waves
calculations.~\cite{Zol03, Jan03,Che94}

Overall, therefore, we see a well-defined trend in the structural and
thermodynamical values shown in Table~\ref{tab:basics}: most quantities
converge rapidly as the basis goes from SZ to DZ to DZP, with DZP
providing an excellent agreement with experiment. Moreover, it appears
that the optimal basis sets, SZ-O and SZP-O, compare very well with
DZP, suggesting that they could be used when computational costs are
an issue. The application of these optimized bases to study the
diffusion of vacancies in GaAs will be reported somewhere
else.~\cite{Elm04b}

\subsection{Defects formation energies in supercell calculations}
\label{sec:formation}

When computing structural and energetic properties of defects using
{\it ab-initio} methods, it is important to ensure that the size of
the basis set is complete enough but also that the simulation cell is
sufficiently large to avoid self-interaction between the defect and
its images.  We have shown previously,~\cite{Elm04} in a study of the
neutral vacancy in silicon, that a supercell of at least 216 atomic
sites can be necessary in order to reduce the elastic and electronic
self-interaction and obtain the right symmetry around the
defect.~\cite{Pro03} As discussed in Sect. \ref{sec:relaxation_As}, we
find a similar behavior for As vacancy; unless indicated, therefore,
we use a 215-atom cell for all our calculations of defects.

In all calculations, this initial 215-atom configuration cell is
randomly distorted, to avoid imposing spurious symmetry in
the fully relaxed defect state.  All atoms are allowed
to relax without any constraint until every force component falls
below 0.04 eV/\AA. The energy minimization takes place at a constant
volume, using the optimal lattice constant obtained with DZP, 5.6\AA\ 
(see Tab.~\ref{tab:basics}), 1\% denser than the experimental value.

The formation energy can be evaluated directly from total energies
obtained from electronic structure calculations. For binary compounds
it is current to use the formalism of Zhang and Northrup~\cite{Zha91}
(see Ref.~\onlinecite{Tor01} for intermediate steps).  The formation energy
of a defect of charge state $q$ is defined as:
\beeq
\label{eq:formation}
E_f = E^{'}_{f} + q(E_V +\mu_e) - {1\over 2}(n_{As} -n_{Ga})\Delta \mu
\eneq
where  $E^{'}_{f} $ is independent of $\Delta \mu$ and $ \mu_e$, and
is represented by
\begin{eqnarray}
E^{'}_{f} &=& E_{tot}(q) -{1\over 2} (n_{As} +n_{Ga})\mu_{GaAs}^{bulk}
-  \nonumber \\ & & -
{1\over 2}(n_{As} -n_{Ga})(\mu_{As}^{bulk} -\mu_{Ga}^{bulk})  
\end{eqnarray}
where $n_{As}$ and $n_{Ga}$ are the number of As and Ga ions present
in the sample, $q$ denotes the net number of electrons or holes
supported by the vacancy, $ \mu_e$ is the electron chemical potential
or the Fermi energy $E_F$, and $E_{V} $ is the energy at the valence
band maximum. Errors in $E_{V} $ due to the finite supercell are
corrected by aligning the vacuum levels of the defective supercell and
undefected supercell~\cite{Zha96}.

If $\Delta \mu$ is defined as the chemical potential difference: 
\beeq
\Delta \mu = (\mu_{As}-\mu_{Ga})-(\mu_{As}^{bulk} -\mu_{Ga}^{bulk}),
\eneq 
the restriction on the chemical potentials becomes: $ 0 \le \mu_e \le
E_g $ and $ -\Delta H \le \Delta \mu \le \Delta H $, where $E_g$ is
the energy gap, and the heat of formation $\Delta H$ of bulk GaAs is
defined as the difference between the chemical potential of bulk As
and bulk Ga crystals and that of bulk GaAs.  This latter quantity
represents the energy necessary to dissociate GaAs into its individual
components: 
\beeq
\label{eq:heat}
\Delta H = \mu_{As}^{bulk} +\mu_{Ga}^{bulk} -\mu_{GaAs}^{bulk}
\eneq

For Ga vacancies of charge $q$ the Eq.~\ref{eq:formation}
reduces to :
\beeq
E_f = E^{'}_{f}   + q(E_V +\mu_e) + {1\over 2}\Delta \mu
\eneq
with $E^{'}_{f} = E_{tot}(q) -{215\over 2} \mu_{GaAs}^{bulk} + {1\over
  2}(\mu_{As}^{bulk} -\mu_{Ga}^{bulk}) $; for As vacancies, it
becomes:
\beeq
E_f = E^{'}_{f} + q(E_V +\mu_e) - {1\over 2}\Delta \mu
\eneq
and $E^{'}_{f}=E_{tot}(q) -{215\over 2} \mu_{GaAs}^{bulk} - {1\over
  2}(\mu_{As}^{bulk}-\mu_{Ga}^{bulk}) $.

\subsection{Computing the ionization energy of charged defects}
\label{sec:ionization}

The concentration of charged defects is controlled by the position of
the Fermi level which is determined by the local concentration of
carriers. Since GaAs is used in a doped state in devices, it is
important to assess the possible charged states of defects.

Charges can affect strongly the formation energy as well as the
structure of a defect, changing the symmetry of the relaxed state and
altering considerably the local electronic properties.  For charged
defects, the effects of finite-size supercell will be even more marked
due to long-ranged nature of the Coulomb interaction; the use a
sufficiently large supercell is therefore even more important.

To account for the electrostatic interaction of periodically arranged
defects of charge $q$ as well as their interaction with the
compensating background, we follow the approximate procedure of Makov
and Payne.~\cite{Mak96} The correction to the total energy of a
charged system is handled by SIESTA, and it consists of a monopole
correction only (~$q^ 2 \alpha/ 2 \varepsilon L $), where $\alpha$ is
the Madelung constant of the simple cubic lattice, $L$ is the
defect-defect distance (16.8 \AA) and $\varepsilon$ is the
experimental static dielectric constant.  The monopole correction is
found to be  0.094 eV, 0.37 eV and 0.84 eV for the charge states
$\pm$1, $\pm$2, and $\pm$3, respectively. The quadrupole correction, which
we evaluated by hand, is proportional to $1/L^3$. For the 215-atom
supercell, it is $2.42 \times  10^{-6} *qQ$ eV (where Q is
the quadrupole moment), and can therefore be neglected (see also
Ref.~\onlinecite{Boc03}).

Because of the limitations of LDA, localized DFT eigenvalues are not
equivalent to the measured electronic levels.  Thus ionization energy
is obtained from the difference between $q_1$ and $q_2$ electron total
energy calculations ($\epsilon(q_2/q_1) = E_{tot}^{q_1} -
E_{tot}^{q2}-(q2-q1)E_V)$, rather than the difference of $q_2$ and
$q_1$ electron eigenvalues of a single calculation. Usually only one
electron is transferred between the electron reservoir and the defect
levels. When two electrons are transferred at the same time the
electron-electron repulsion is compensated by a relaxation of the
structure around the defect that arises from a strong electron-phonon
coupling.  This so-called {\it negative-U} effect is found when the
ionization level $\epsilon(q-1/q)$ appears above $\epsilon(q/q+1)$,
thus a direct transition $(q-1/q+1)$ is energetically more
favorable.~\cite{And75}

\section{Converging defect formation and ionization energies}
\label{sec:effects}
In this section, we study the effects of the basis set and the
$k$-sampling and the simulation cell on the formation energy and the
relaxed geometry of neutral and charged Ga and As vacancies.

\subsection{Local basis set effect}
\label{sec:basis}

We first study the effect of the choice of the local basis set in the
defects.  In order to cancel out all other effects, we use supercells
of 215 atomic sites and a Mokhorst-Pack grid~\cite{Mon76} of
2$\times$2$\times$2 corresponding to a density of 0.03\AA$^{-1}$ of
k-points.

Cohesive energies as well as bulk moduli studied in
Section~\ref{sec:total} for the different basis sets used show that
atomic bonding is strengthened progressively as we go from SZ basis to
DZP.

Structural relaxation is directly related to the interatomic forces
acting on the atoms around the vacancy and on the strength of atomic
bonding. Atoms around the vacancy form initially an ideal tetrahedron
with six equal distances labeled $d_1-d_6$ with tetrahedral
symmetry $T_d$. After the full relaxation (see Sect.~\ref{sec:total}),
distances and angles can be altered and the symmetry is either conserved
or broken.

Relaxations around the vacancies are given for different charge states
in Tables~\ref{tab:relax_Ga} and \ref{tab:relax_As} for Ga and As
vacancies respectively. In both cases the corresponding formation
energy is reported as well as the relative change in the volume of the
tetrahedron with respect to the ideal one.~\cite{Pro03, Tor01}

Due to the finite precision in the relaxation, there is some
imprecision in the identification of the defect symmetry. Here, if the
highest relative difference between two bonds is lower than 1 \%
(equivalent to a precision of 0.04 \AA), the structure is assigned to
highest symmetry group. The last column lists the symmetry groups for
the different defects in the DZP basis set.  Unless specified, the
symmetry group for all bases is the same as that of DZP.

\subsubsection{$V_{Ga}$}
\label{sec:effect_Ga}

\begin{table*}
\caption{Convergence of the formation energy $E^{'}_f$ in eV with
  respect to the basis set for the Ga vacancy. Relaxation around
  the vacancy are given in \% compared to the ideal tetrahedral 
  distance between {\bf As} nearest neighbors. The distances are
  labeled $d_1-d_6$, the negative sign indicates an inward relaxation. The
tetrahedron volume change is also given in \% of the
  ideal volume ($\Delta V = 100 *(V-V_0)/V_0$). The last column
  displays the symmetry group of the defect (see the text for more
  details).  } 
\label{tab:relax_Ga}
\begin{ruledtabular}
\begin{tabular} {ccccccccccc}
& & \multicolumn{6}{c}{Distances in \%  } \\
 \cline{3-8}
Basis & $E^{'}_f$(eV)  &$d_1$ &$d_2$ &$d_3$ &$d_4$ &$d_5$ &$d_6$
 &$\Delta V$ &Symmetry \\ \\ 
 \multicolumn{10}{c}{$V_{Ga}^0$} \\ 
\\
\hline
 SZ&2.7 &-19.3 &-19.2 &-19.3  &-19.3  &-19.2  &-19.3  &-47.3  \\
 DZ&2.8 &-14.5  &-14.4  &-14.5  &-14.4  &-14.5  &-14.4  &-37.4  \\
 {\bf DZP} &{\bf2.9} &{\bf-13.5}  &{\bf-13.4}  &{\bf-13.5}
 &{\bf-13.4}  &{\bf-13.4}  &{\bf-13.5}  &{\bf-35.2} &{\bf$T_d$}\\ 
 
 \\
 \multicolumn{10}{c}{$V_{Ga}^{-1}$ }\\
 \\
  \hline
 SZ&2.9 &-19.0 &-19.1 &-19.0 &-19.0 &-19.0 &-19.0 &-46.9 \\
DZ& 2.9&-14.7 &-14.7 &-14.7 &-14.7 &-14.7 &-14.7 &-37.9\\
{\bf DZP}&{\bf3.0} &{\bf-14.2} &{\bf-14.2} &{\bf-14.2} &{\bf-14.2}
 &{\bf-14.2} &{\bf-14.2} &{\bf-36.9} &{\bf$T_d$}\\

\\
 \multicolumn{10}{c}{$V_{Ga}^{-2}$} \\
 \\
 \hline
SZ&3.4 &-19.5 &-19.5 &-19.5 &-19.5 &-19.5 &-19.5 &-47.8 \\
 DZ&3.2 &-14.4 &-14.4 &-14.4 &-14.4 &-14.4 &-14.4 &-37.3\\
{\bf DZP}&{\bf3.4}  &{\bf-14.0} &{\bf-14.1} &{\bf-14.0} &{\bf-14.0} &{\bf-14.1} &{\bf-14.0} &{\bf-36.5} &{\bf$T_d$}\\

\\
 \multicolumn{10}{c}{ $V_{Ga}^{-3}$ }\\
 \\
 \hline
SZ&4.1&-19.7 &-19.8 &-19.7 &-19.7 &-19.8 &-19.7 &-48.3 \\
DZ&3.8 &-15.1 &-15.2 &-15.1 &-15.1 &-15.2 &-15.1 &-38.9\\
{\bf DZP} &{\bf3.9} &{\bf-14.6} &{\bf-14.6} &{\bf-14.6} &{\bf-14.6} &{\bf-14.7} &{\bf-14.6} &{\bf-37.7} &{\bf$T_d$}\\
\end{tabular}
\end{ruledtabular}
\end{table*}

The Ga vacancy maintains the same symmetry for all charge states
irrespective of the basis set used. The structural relaxation is most
important for the smallest basis set, SZ, decreasing progressively by
about one third as the number of bases is increased, but the $T_d$
symmetry is maintained in all cases, with the atoms moving inwards
systematically. Moreover, for each basis set the degree of structural
relaxation, is almost independent of the charge state, contracting
very slightly, by about 2 \%, from the neutral to the -3 charge state,
for all basis sets.

As can be seen in Table~\ref{tab:relax_Ga}, the formation energy is
also rather well converged with the minimal basis set (SZ), by
comparison with the more accurate DZP: the difference between the two
bases is at most 0.2 eV.

Heavily charged defects like $V_{Ga}^{-3}$, which is the most likely
charge state in a heavily doped material, are more sensitive to the
completeness of the basis set. This effect explains the fact that the
formation energy decreases with an improved basis, contrary to the
other defects. 

Because of its technological importance, we must ensure that the
orbital overlap around the vacant site is sufficient to accommodate
the extra electrons in $V_{Ga}^{-3}$. We can do so by placing a {\it
  ghost} Ga atom at the defect site.  For that purpose, we generate a
set of orbitals and place them on the crystalline site, without adding
any pseudo-potential or extra electrons. The system is then relaxed
using the same convergence criterion as before. This ghost atom does
not have any significant effect on the total energy of the defect when the DZ
and the DZP basis sets are used. In contrast, SZ basis total energies
are corrected by 0.34 eV, suggesting that the SZ orbital are too
short.  This additional set of orbitals is sufficient to correct for
the overestimated formation energy with SZ basis, decreasing its
value from 4.06 to 3.72 eV, following the general trend observed for
other charge states (see table~\ref{tab:relax_Ga}).

\subsubsection{$V_{As}$}
\label{sec:effect_As}

The situation is very different for the As vacancy: the local symmetry
is broken for most charge states and the completeness of the basis set
impacts strongly on the reconstruction around the defect.  Except
for the positively charged vacancy, the bonds are stretched
considerably to form pairs, leading to volume deformation by as much
as 60\%. Because of this strong deformation, we must relax the
criterion on the symmetry.  To allow the reader to judge the
classification, we write the distances $d_1-d_6$ in ascendent order in
the last column of table~\ref{tab:relax_As}.

\begin{table*}
\caption{Convergence of the formation energy $E^{'}_f$ in eV with
  respect to the basis set for the As vacancy. The relaxation around
  the vacancy is given in \% compared to the ideal tetrahedral
  distance between {\bf Ga} nearest neighbors. The distances are
  labeled $b_1-b_6$, the negative sign indicates an inward relaxation. The
  volume change around the vacancy  is also given in \% of the
  ideal volume ($\Delta V = 100 *(V-V_0)/V_0$). The last column
  displays the symmetry group of the defect (see the text for more
  details). }
\label{tab:relax_As}
\begin{ruledtabular}
\begin{tabular} {ccccccccccc}
& &\multicolumn{6}{c}{Distances in \%  } \\
 \cline{3-8}
Basis & $E^{'}_f$(eV) &$d_1$ &$d_2$ &$d_3$ &$d_4$ &$d_5$ &$d_6$ &$\Delta V$ &Symmetry \\ \\

 \multicolumn{10}{c}{$V_{As}^{+1}$ }\\
 \\
 \hline
SZ&2.2 &-14.5 &-14.9 &-15.1  &-15.7 &-15.9  &-16.6       &-39.5  \\
DZ& 2.7 &-7.9 &-8.0 &-8.1 &-8.4    &-8.4 &-8.7   &-22.7 \\
{\bf DZP}&{\bf2.8}  &{\bf-5.6} &{\bf-5.9} &{\bf-6.1} &{\bf-6.3}  &{\bf-6.2}   &{\bf-6.4}        &{\bf-17.1} &{\bf~$T_d$}\\  

 \\
 \multicolumn{10}{c}{$V_{As}^0$ }\\
 \\
 \hline
SZ& 2.4 &-13.8 &-14.3 &-15.2     &-15.2 &-31.3       &-31.7     &-53.4  &$C_{1h}$\\
 DZ&3.1&-12.2 &-12.2 &-12.7 &-13.0   &-26.9      &-27.0     &-46.7  &$\sim D_{2d}$\\
{\bf DZP} &{\bf3.2} &{\bf-11.1} &{\bf-11.2}  &{\bf-11.5} &{\bf-11.6}  &{\bf-24.6}      &{\bf-24.5}    &{\bf-43.0} &{\bf $D_{2d}$} \\

 \\
 \multicolumn{10}{c}{$V_{As}^{-1}$  }\\
\\
\hline
SZ &2.5 &-14.3 &-14.7 &-15.3 &-15.4  &-33.3      &-33.5    &-55.6 \\
DZ&3.2 &-12.5  &-12.7 &-13.2 &-13.4  &-31.8    &-32.0    &-52.4  \\
{\bf DZP}&{\bf 3.3 } &{\bf -11.6} &{\bf-11.9} &{\bf-12.5} &{\bf-12.6}  &{\bf-31.2}      &{\bf-31.5}    &{\bf-51.2} &{\bf$\sim D_{2d}$} \\

 \\
 \multicolumn{10}{c}{$V_{As}^{-2}$  }\\
 \\
 \hline
SZ &3.8 &-17.8 &-18.5 &-18.5 &-19.3  &-33.6      &-33.7    &-58.6  &$\sim C_{3v}$\\
DZ &4.4 &-16.5 &-17.1 &-17.5 &-18.2  &-32.0      &-32.3    &-56.2 &$\sim C_{3v}$\\
{\bf DZP}&{\bf4.6 } &{\bf-16.5} &{\bf-16.8} &{\bf-17.0}  &{\bf-17.6}  &{\bf-31.6}     &{\bf-31.7}    &{\bf-55.4}  &{\bf$C_{3v}/ D_{2d}$}\\

 \\
 \multicolumn{10}{c}{$V_{As}^{-3}$ }\\
\\
\hline
SZ &4.9 &-18.4  &-18.4 &-31.4  &-31.6    &-31.7  &-31.8  &-65.5  \\
DZ & 5.4 &-18.3  &-18.2 &-30.4  &-30.6   &-30.7  &-31.0  &-63.8 \\
{\bf DZP }&{\bf5.9} &{\bf-18.1} &{\bf-18.2 } &{\bf-30.4 } &{\bf-30.2}  &{\bf-30.2}&{\bf-30.2}&{\bf-62.9 }&{\bf$D_{2d}$-resonant}\\

\end{tabular}
\end{ruledtabular}
\end{table*}

As a general trend, the SZ basis is less efficient for As than for Ga
vacancies: the formation energy is underestimated by as much as 0.6 eV
(27\%) for charge +1 and 1.0 eV (17\%) for charges -3 as compared with
DZP. The underestimation drops to 2-8 \% with DZ, a considerable
improvement, for all charge states. The improvement in formation
energy can be directly correlated with the prediction quality of the
local relaxation. For example, while $T_d$ symmetry is conserved for
the three basis sets for $V_{As}^{+1}$, the change in the
volume around the $V_{As}^{+1}$ is highly overestimated by 135 \%
using SZ basis compared to the DZP results. The overestimation drop to
35 \% with DZ, leading to an error of less than 0.1 eV compared with
DZP.

As more electrons are added to the defect level this trend tends to
diminish, the symmetry and the relaxation of the defect can be
described with reasonable accuracy using the DZ basis.  Applied to the
singly negative As vacancy $V_{As}^{-1}$, the inclusion of the ghost
atom at the vacant site has a smaller impact on the energy level and
relaxation using SZ than for $V_{Ga}^{-3}$, and next to none with DZ
and DZP. For the minimal basis set, the correction is negligible and
accounts only for 0.067 eV. 

\subsubsection{Convergence of the ionization energy with the basis sets}
\label{sec:ioniz_cov}

As defined in section~\ref{sec:ionization}, ionization energies are
taken as the difference between the total energy in different charge
states. Unlike formation energies: the errors introduced for the
ionization energies are considerably reduced errors coming from LDA
are canceled out and those coming from different
chemical potentials are eliminated. The only remaining error are those
coming from the basis set convergence, the evaluation of the valence
band edge energy and the Madelung correction.

\begin{figure}
\centerline{\includegraphics[width=5cm]{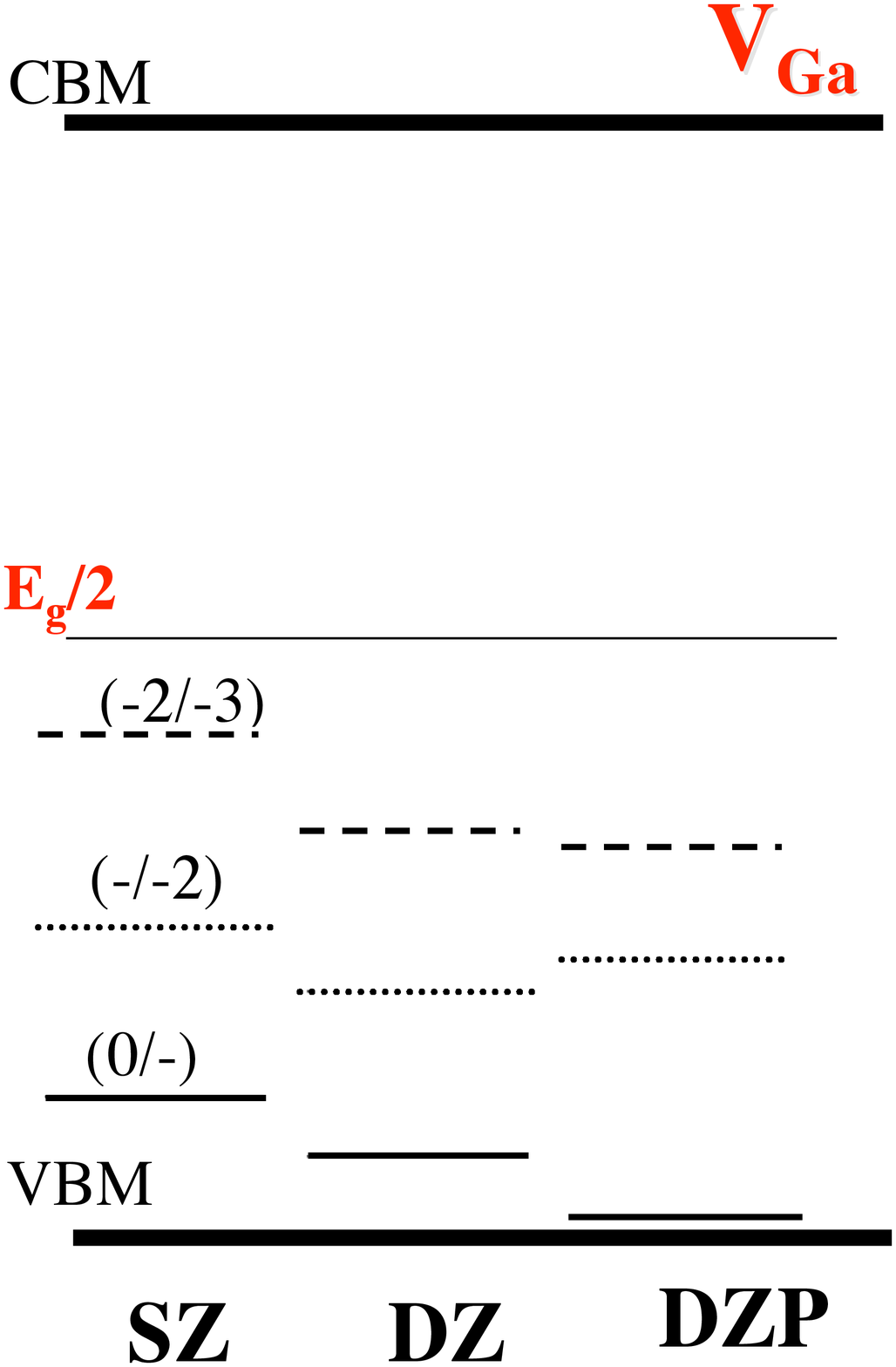}\includegraphics[width=5cm]{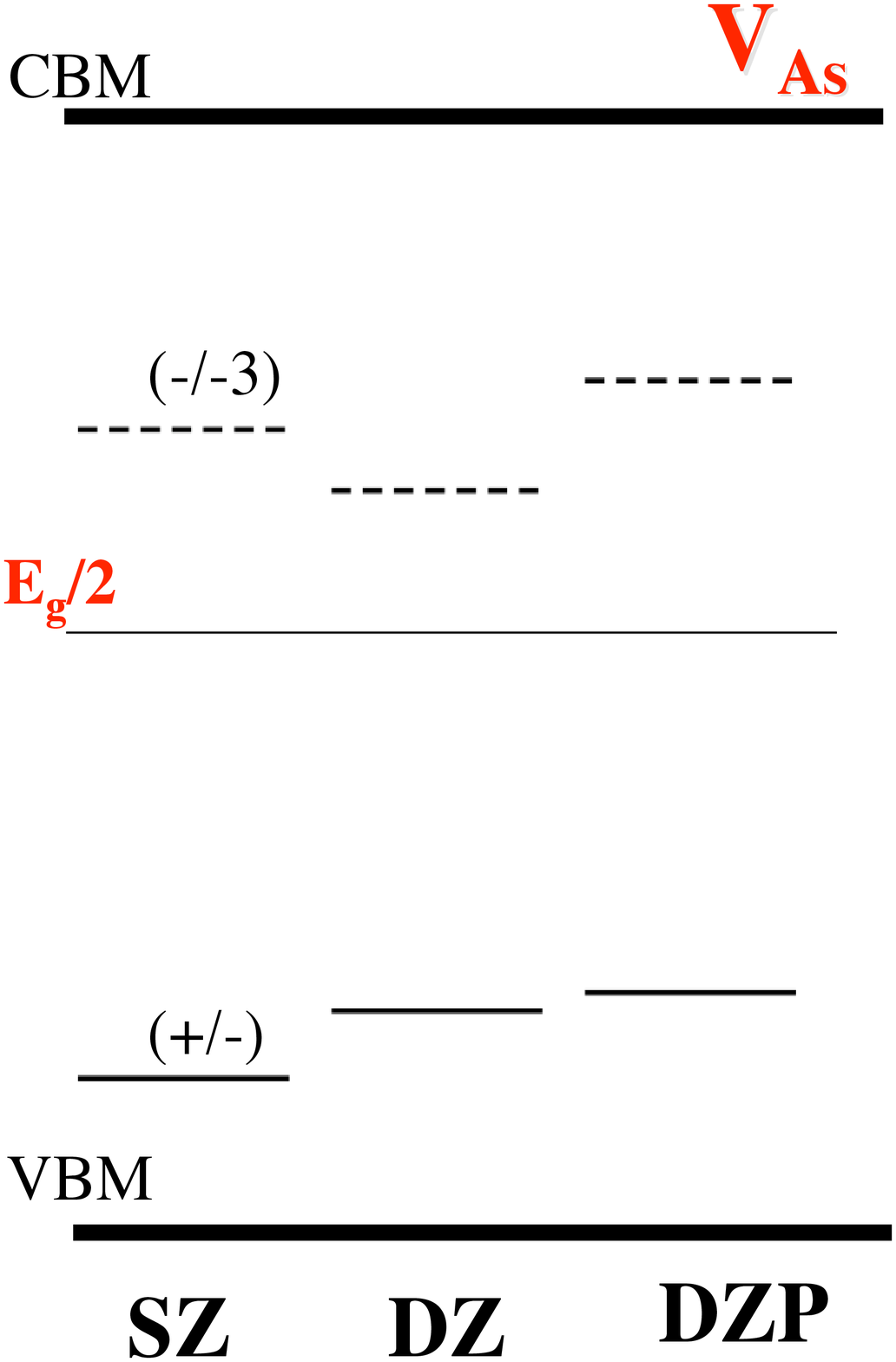}}
\caption{Schematic representation of the convergence of ionization
  energies as function of the basis set. Due to the underestimation of
  the gap, as a consequence of LDA, it is usual to align the
  conduction band maximum (CBM) with the experimental value, at 1.54 eV
  from the valence band maximum (VBM).  The left panel displays the
  ionization energies as function of the basis set for the Ga
  vacancy that are all all below the experimental mid gap. The right
  panel shows the two ionization levels located below and above the
  mid gap and their convergence with respect to the different basis
  sets used. Refer to the text for the basis set description.}

\label{fig:levels}
\end{figure}

Figure~\ref{fig:levels} displays the convergence of the ionization
energies as a function of the basis set used for $V_{As}$ and
$V_{Ga}$. For both types of defects, the preliminary results obtained
using SZ basis set give a rough estimate of the location of ionization
energies in the band gap.  These energies converge with increasing
basis sets, but slower than the formation energies.  For $V_{Ga}$,
while the formation energies are reasonably converged already with the
minimal basis set, the ionization levels found using the SZ basis are
noticeably overestimated compared to DZP. The levels are located at
0.21, 0.43 and 0.69 eV for SZ, while for DZ it changes for 0.13, 0.35
and 0.57 eV then finally converges to 0.05, 0.4 and 0.55 eV for DZP.
For $V_{As}$, we found in section~\ref{sec:effect_As} that formation
energies are underestimated by less than 1eV with SZ for most charge
states. Once ionization energies are calculated, errors on the
formation energies coming from the relaxation around the defect cancel
out since defects in all charge states suffer from this effect.

SZ basis gives a correct qualitative description of the nature of the
electronic transition (double negative-U effect, to be discussed in
section~\ref{sec:relaxation_As}). The location of the levels in the
band gap as well as the distance between them are also reasonably
converged.  Ionization energies with SZ are slightly underestimated
compared to the DZP ionization levels. Levels are at 0.13 and 1.19 eV
for SZ , 0.24 and 1.09 for DZ and finally 0.27 and 1.27 for DZP.

By studying the effect of the choice of the basis set on the
structural relaxation, the formation and the ionization energies, we
conclude that the SZ basis is significantly less efficient when the
local symmetry is broken. In these cases, the use if a second radial
function (DZ) is necessary to obtain reasonable numbers. Moreover,
both DZ and DZP are complete enough to represent the properties
associated with a defect without the need for a ghost atom.

\subsection{K-points effects}

The effects of Brillouin zone sampling are studied by comparing the
formation energies and the relaxation volume of both As and Ga
vacancies for all charge states. We consider two k-point sampling, a
$\Gamma$ point sampling and a $2\times2\times2$ Monkhorst-Pack
mesh~\cite{Mon76}, corresponding, for the 215-atom cell used here, 
to a density of 0.06\AA$^{-1}$ and 0.03\AA$^{-1}$, respectively. To
isolate sampling effects, all calculations are done using the DZP
basis set. Results are shown in Table~\ref{tab:k_points}.

\begin{figure}
\centerline{\rotatebox{-90}{\includegraphics[width=11cm]{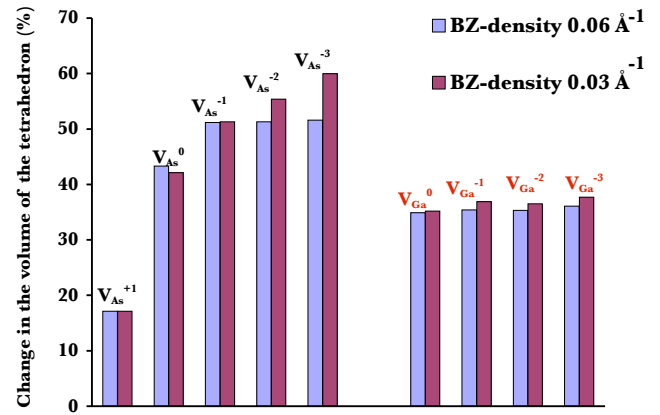}}}
\vspace{-4cm}
\caption{(Color online) Histogram of the change in the volume of the
  relaxed tetrahedron formed by  atoms surrounding the vacancy in \%
  of the ideal one for two densities of k-points in the Brillouin zone
  (refer to the text for details). For comparison,  data for  Ga
  and As vacancies in different charge states are plotted in the
  same figure.} 
\label{fig:volume}
\end{figure}

The use of $\Gamma$ point only gives formation energies that are
reasonably converged for Ga vacancies. Similarly, the bond lengths
around the vacancy and the relaxation volume (Fig.~\ref{fig:volume})
are near those of the $2\times2\times2$, preserving the $T_d$
relaxation symmetry for all charge states. 

The use of $\Gamma$-point only produces less reliable results
in the case of As vacancies. Formations energies are well converged
but the relaxation symmetry around the defect is not correctly
predicted for all charge states.  With $\Gamma$-point, the distance
between the atoms forming the tetrahedron as well as its associated
volume are already converged for charge stated +1, 0, and -1, as shown
in Fig.~\ref{fig:volume}.

For $V_{As}^{-2}$, $\Gamma$-point sampling overestimates the volume
contraction around the defect by 7.5\%. The long bonds in the pairing
mode associated with the $D_{2d}$ structure are shortened from 3.46 \AA\ 
for this sampling to 3.26 \AA\, for $2\times2\times2$ sampling.  The
k-point sampling effect becomes even stronger for the highly charged
$V_{As}^{-3}$. The resonant bond geometry present with high-density
sampling (discussed further in Sect. ~\ref{sec:relaxation_As}) is not
found with the $\Gamma$-point sampling and the defects relaxes into
the usual pairing mode with two short bonds and four long bonds.

Comparing the two samplings for $V_{As}^{-3}$, we find that: (1) with
DZP + $2^3$ mesh, the resonant-bond configuration has the lowest
energy; (2) the pairing mode configuration is unstable using DZP +
$2^3$ mesh, starting from this configuration, the defect relaxes back
into the resonant-bond state; (3) for $\Gamma$-point sampling the
resonant bond configuration is found to be metastable but the pairing
mode is favored with an energy difference of 0.6 eV.

Overall, the use of $\Gamma$-point sampling gives a reasonable
description of the energetic and structural properties of the Ga
vacancy as well as most charge states for As. It fails, however, for
heavily charged defects, when many electrons are involved in the
bonding.

\begin{table}

\caption{Convergence of the formation energies  $E_{f}^{'}$ (in eV)
  for Ga and As vacancies with respect to the Brillouin zone
  sampling. } 

\label{tab:k_points}
\begin{ruledtabular}
\begin{tabular} {lcccccc}

&+1 &Neutral &-1 &-2 &-3   \\   
&\multicolumn{5}{c}{$E_{f}^{'}(V_{Ga})$} \\     
\cline{3-6}
{$\Gamma$} &&2.51 &2.73 &3.17 &3.81 \\
{$2\times2\times2$ } &&2.94 &3.00 &3.40 &3.94\\ &&15\% &9\%  &7\%  &3\% \\
\\
 &\multicolumn{5}{c}{$E_{f}^{'}(V_{As})$} \\    
  \cline{2-6}
{$\Gamma$} &2.73 &3.21  &3.52 &4.70 &6.08        \\      
{$2\times2\times2$ }  &2.79&3.25  &3.33 &4.52 &5.86    \\              &-2\% &-1\% &6\% &4\% &4\%     \\
\end{tabular}
\end{ruledtabular}
\end{table}

\subsection{Size effects}

Size effects on defect formation energies have been widely discussed
for a number of systems including the silicon vacancy (see
Ref.~\onlinecite{Elm04} and references therein) and GaAs.~\cite{Sch02}
Size effects are found to be strong for cubic supercells smaller than
216$\pm$1 for both materials and becomes negligible for larger
systems. For this reason, we will study in this section size effects
on charged Ga and As vacancies for supercells of 216 ionic sites and
smaller.

To characterize the size effects, we consider the dominant charge
state for each vacancy type, $V_{As}^{-1}$ and $V_{Ga}^{-3}$, with DZP
basis set. We simulate two cell-size with the same k-point density: a
63-atom unit cell with a 3$\times$3$\times$3 sampling and a 215-atom
cell with $2\times2\times2$ sampling.~\cite{Sch02}

We find a formation energy of 3.83 eV compared to 3.33 eV obtained for
the 215 cell for $V_{As}^{-1}$. In this case size effects are strong
and overestimate the formation energy by 0.5 eV. More important, the
symmetry of the relaxed defect is different for the two sizes: in the
63-atom cell, the defects relaxes to a $C_{2v}$ symmetry while the
defects 215-atom cell adopts a $D_{2d}$ symmetry
(Table~\ref{tab:relax_As}).  We check that the 63-atom cell is not
caught into a metastable state by relaxing the defect starting in a
$D_{2d}$ symmetric state. After full relaxation, the cell relaxes into
a $C_{2v}$, confirming that this is the lowest-energy symmetry for
this cell. The different symmetry also impacts the change in the volume
surrounding the defect. As a consequence of the strong defect-defect
interaction in the smaller cell, the structural relaxation is hindered
and the volume decreases by 43.8\% compared with 51.2\% for the
215-atom cell.

As could be expected from the previous sections, size effects are
less important for $V_{Ga}^{-3}$. In particular, the initial
tetrahedral symmetry ($T_d$), with atoms equidistant, is maintained
around the defect for the relaxed 215-atom supercell. The same
symmetry is found in the 63-atom supercell.  Intercell defect-defect
interaction rigidify the lattice, however, and the change is volume is
only 34.8 \% for the 63-atom cell compared to 37.7 \% for the larger
supercell.  As with $V_{As}^{-1}$, the formation energy is
overestimated with the small cell: $E^{63}_f = 4.24$ eV while
$E^{215}_f =3.94$ eV).  This difference is considerable as it is on
the order of the ionization energies.

In this section, we have studied in details the effects of the choice
of basis set, k-point sampling and simulation cell on the properties
of charged defects. In summary, we find that: (1) The DZP basis set is
well-converged and ensures reliable results for all charge states.
For a number of charged states, it is also possible to use cheaper
optimized basis sets for a similar accuracy. This is not always the
case, however, and the applicability of these basis sets must be
evaluated on a case by case basis.  (2) SZ is less efficient than DZP
for defects where symmetry is broken, but it gives a satisfactory
estimation of the location of ionization levels in the band gap. (3)
For supercell of 216 atoms or more, the density of k-points has only
a minor effect on the defect relaxation. The use of the $\Gamma$ point
only gives a relaxation and a symmetry that are satisfactory, in most
cases. However, this reduced sampling must be used with care for
highly charged defects such as $V_{As}^{-3}$. (4) The errors arising form
size effects are much more important than those coming from the
density of k-points.  In particular, size effects can be the source of
errors in estimating the ionization energies, especially when the
transition from a charge state to another induces breaking of the
symmetry.

\section{Results and discussion}
\label{sec:results}

We discuss here the results reported in  the previous sections; we
concentrate on the highly converged results for the Ga and As
vacancies, obtained using the DZP basis set using a
215-supercell and a Monkhorst-pack grid of $2\times2\times2$ in the
reciprocal space. We invite the reader to refer to the DZP results
reported, in bold, in Tables~\ref{tab:relax_Ga} and
~\ref{tab:relax_As}. We first deal with the stochiometric case where
$n_{Ga} =n_{As}$ ($\Delta \mu$ =0), then we study the formation
energies for each kind of defects  separately under ideal
growing conditions as a function of the doping level. Next, the
dominant vacancy type defects in real GaAs crystals are identified by
taking into account growing conditions. The ionic chemical potentials
varies from As-rich conditions ($\Delta \mu= -\Delta H$) to Ga-rich
conditions ($\Delta \mu= +\Delta H$) as the Fermi level is changed
progressively.
 
\subsection{Gallium vacancies}
\label{sec:Ga}
\begin{table}
\caption{ Comparison between  ionization energies (measured from the
  valence band edge) of  the Ga vacancy in GaAs. Results are grouped
  following the three classes discussed in the text.   } 
\label{tab:comp_Ga}
\begin{ruledtabular}
\begin{tabular} {llllll}
&\multicolumn{5}{c}{Ionization levels (eV)} \\  
\cline{2-6}
&&&&\multicolumn{2}{c}{Negative U}\\
\cline{5-6}
Authors  &0/-1 &-1/-2 &-2/-3 &+1/-1 &-1/-3\\
\hline\\
Seong and Lewis ~\cite{Seo95}&&&&0.035 &0.078 \\
\\
Northrup and Zhang~\cite{Nor93}&0.19 &0.2 & 0.32 \\
P\"oykk\"o et al~\cite{Poy96} &0.11 &0.22  &0.33\\
Schick et  al.~\cite{Sch02}&0.09 &0.13 &0.2 \\
Janoti et al. ~\cite{Jan03} &0.13 &0.15 &0.18 \\
\\
Jansen and Sankey ~\cite{Jan89} &0.1&0.35 &0.50 \\
Baraff and Schl\"uter~\cite{Bar85}&0.2 &0.5 &0.7\\
Cheong and Chang~\cite{Che94}&&0.49 &0.69 \\
Gorczyca et al~\cite{Gor02}&0.39 &0.52 &0.78 \\
This work &0.05 &0.4 &0.55 \\
\end{tabular}
\end{ruledtabular}
\end{table}

\subsubsection{Relaxation geometry }

\begin{figure}
\centerline{\resizebox{10cm}{!}{\includegraphics[width=3cm]{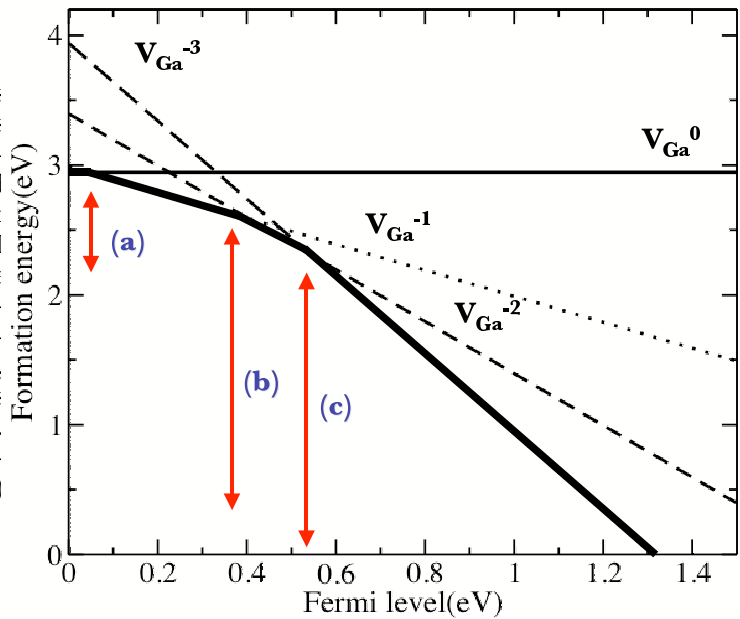}}}
\caption{(Color online) Formation energies as function of Fermi
  level in  various charge states of Ga vacancies at 0K. The
  Fermi level is defined by reference to the valence band maximum. Ionization
  levels are defined as the intersection between the formation
  energies of different defects. Defect with the lowest formation
  energy is dominant.  Arrows point to the location of the ionization
  levels labeled (a) for (0/-), (b) for (-/-2) and (c) for (-2/-3) } 
\label{fig:levels_Ga}
\end{figure}

Table~\ref{tab:relax_Ga} describes the fully relaxed geometry of the
defect. The structural deformation obtained is well localized around
the vacant site; the magnitude of the relaxations is listed for the
nearest neighbor As atoms in all relevant charge states, reaching 15\%
of the bulk bond distances. The tetrahedral symmetry $T_d$ is always
conserved for this defect, irrespective of the charge state: only the
breathing mode  matters here.  The As dangling bonds do not form pairs
in any charge state, but the back bonds formed with Ga atoms are
clearly weakened; this is in agreement with the  observation
that the pairing mode is generically not energetically favorable for
cation vacancies (Ga).~\cite{Cha03}  We also observe that
all As atoms relax inward but the amount of the relaxation does not
increase as more electrons are added to the vacancy levels but remains
stable (13.5-14.5\%). Our results are in good agreement with the
results of Laasonen et al.~\cite{Laa92}, and Seong and
Lewis~\cite{Seo95}, who found a systematic inward relaxation with
tetrahedral symmetry for Ga vacancies in (0,-1,-2) charge state
studied using an empirical tight-binding potential.

\subsubsection{Energetics}

Experimentally, the Ga vacancy is found to exist in the 0,$-$,$-2$,$-3$
charge states. Until recently, the preferred charge state for the Ga
vacancy in GaAs was the subject of a hot debate: most {\it ab-initio}
calculations~\cite{Seo95, Poy96, Nor93,Che94} find that where GaAs in
either semi-insulating or n-type and Fermi energy is away from the
valence band edge, the gallium vacancy is in the triply negative
charge, while diffusion experiments suggest a charge of $-$2 or
$-1$.~\cite{Bra99}
 
Using positron annihilation to determine the Gibbs free energy of
formation for Ga vacancies in GaAs, Gebauer {\it et al.}~\cite{Geb03}
could finally resolve this debate, giving a quantitative estimation of
the formation enthalpy for $V_{Ga}^{0}$ and $V_{Ga}^{-3}$. The vacancy
concentration is directly probed with positron annihilation in Te
doped GaAs as function of doping concentration, temperature and
chemical potential.  Our estimate of the formation energies reported
in the first column of Table~\ref{tab:relax_Ga} are in good agreement
with recent experimental and theoretical data. For the neutral
vacancy, we get $E_{f}^{'}$($V_{Ga}^{0}$)= 2.94 eV, a values that
agrees with $E_{f}^{'}$($V_{Ga}^{0}$)=2.8 eV from Bockstedte and
Scheffler~\cite{Boc96} (earlier
first-principals calculations~\cite{Nor93} predict 3.5 eV) and the
experimental results of Gebauer {\it et al.},~\cite{Geb03}
$H_{f}^{'}$($V_{Ga}^{0}$)=3.2$\pm$ 0.5 eV.
A formation enthalpy of 1.8$\pm$0.5 eV was also measured by Mitev {\it
  et al.}~\cite{Mit98} using inter-diffusion experiment on AlGaAs/GaAs
heterostructure. However, the charge state of the associated defect is
unknown, rendering the comparison with our results difficult.

For the triply negative charge state, we find that
$E_{f}^{'}$($V_{Ga}^{-3}) = 3.9$~eV. This value compares well with
recent experiments~\cite{Geb03} which give $H_{f}^{'}$($V_{Ga}^{-3}) =
3.6$~eV as well as with recent theoretical study by Janoti {\it et al.}
~\cite{Jan03} ( 3.6 $\pm$ 0.2 eV ).

Considering the stochiometric case where $n_{Ga} =n_{As}$ ($\Delta
\mu$ =0), we can study the formation energies of GaAs under ideal
growing conditions as a function of the doping level.
Figure~\ref{fig:levels_Ga} displays the formation energies as function
of Fermi energy for various charge states of Ga vacancies at 0K.  As
$E_f$ depends linearly on the electronic chemical potential,
$\mu_{e}$, the slope of each of the lines represents the net charge
for the system $q$. Intersections determine the ionization levels
where one electron is transferred from the electron reservoir to the
defect level. We can see that at each transition only one electron is
transferred at a time. Moreover, the ionization levels labeled a, b, and
c favor the stability of the $-3$ charge states for intrinsic and n-type
GaAs.

In order to compare with earlier theoretical work, we summarize the
results into three types of behavior:
(1) Sole among all calculations,  Seong et Lewis~\cite{Seo95} find a
negative-U effect for the Ga vacancy using TB-MD method, the
transition levels identified are very shallow and favor the triply
negative state in almost the entire range of the Fermi level.
(2) The second category of levels reported in
Ref.~\onlinecite{Poy96,Nor93, Sch02, Jan03} are shallow and lie well
below the mid gap with no negative-U effect detected.
(3) Defects level can also lie deeper below the mid gap. This is the
case for levels a, b and c in  Fig.~\ref{fig:levels_Ga}, which compare
well with the results of Cheong and Chang~\cite{Che94},
Baraff and Schl\"uter~\cite{Bar85} and a recent study form Gorczyca et
al~\cite{Gor02} where the $-3$w charge of $V_{Ga}$ is relevant only if
the Fermi energy is above 0.55 eV.

Unfortunately, ionization levels cannot be directly measured
experimentally, only their sum is obtainable.  Nevertheless,
experimental values obtained by electron irradiation of
GaAs~\cite{Jia92} support the assignment of deep-lying levels in GaAs,
in agreement with the third category. Gebauer {\it et
  al.}~\cite{Geb03} confirm 
this assignment using a model to fit their
experimental data in order to identify the charge state of the vacancy
in GaAs from the location of the ionization levels. This calculation
shows that the -3 charge state is the most stable charge state if deep
ionization energies coming from a calculation on unrelaxed Ga
vacancies of Baraff and Schl\"uter,~\cite{Bar85} which are in
agreement with our more precise calculations.

\begin{table}
\caption{Comparison between  ionization energies (measured from the
  valence band edge) 
  for the As vacancy in GaAs. The data are grouped according to the
  two categories discussed in the text.  } 
\label{tab:comp_As}
\begin{ruledtabular}
\begin{tabular} {llllll}
&\multicolumn{5}{c}{Ionization levels (eV)} \\  
\cline{2-6}
&&&&\multicolumn{2}{c}{Negative U}\\
\cline{5-6}
Authors &+2/+1 &+1/0 &0/-1 &+1/-1 &-1/-3 \\
\hline\\
Seong and Lewis ~\cite{Seo95}&&1.41 &1.54 \\
Jansen and Sankey ~\cite{Jan89} &&1.3&1.40\\
\\
Cheong and Chang~\cite{Che94}&&&&0.785 \\
P\"oykk\"o et al~\cite{Poy96} &&&&0.86\\
This work &&&&0.27 &1.27 \\
\end{tabular}
\end{ruledtabular}
\end{table}

\begin{figure}
\centerline{\includegraphics[width=10cm]{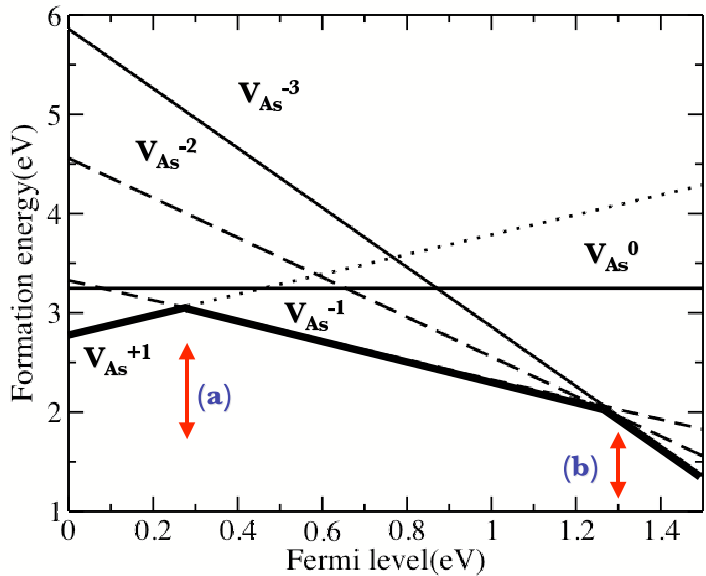}}
\caption{(Color online) formation energies as a function of the Fermi
  energy of various charge states of As vacancies at 0K. The Fermi
  level is calculated with respect to the valence band maximum. Arrows
  point to the location of the ionization levels labeled (a) for
  ($+/-$), (b) for ($-/-3$)}
\label{fig:levels_As}
\end{figure}

\subsection{Arsenic vacancies}
\label{sec:As}

\subsubsection{Relaxation geometry }
\label{sec:relaxation_As}

As vacancies in various charge states have been studied by a number of
authors~\cite{Seo95,Laa92, Fen97} taking into consideration ionic
relaxation and relaxation geometry. Except for $V_{As}^{+1}$,
breathing mode displacement breaks the vacancy local
symmetry.~\cite{Seo95,Laa92} Contrary to Ga vacancy, the volume of the
As vacancy increases as electrons are added and shrinks when electrons
are removed.

Our results confirm part of these findings. From
Table~\ref{tab:relax_As}, the volume of the tetrahedron shrinks from
-17\% to -60\% of the initial volume as electron are added to the
vacancy. However, all charge states, even positive ones, display
inward relaxation with respect to the unrelaxed volume. 

Comparing the bond lengths obtained after a full relaxation of the
structure (Table~\ref{tab:relax_As}), we find that for the positively
charged vacancy ({$V_{As}^{+1}$}) there is no electron in the
localized states and all four atoms relax inward by about -6\%
conserving the tetrahedral symmetry, with no Jahn-Teller distortion.
Although this inwards relaxation is more important, in absolute value,
than the outward relaxation reported in references\cite{Laa92,Seo95}, they
agree in term of the conservation of the symmetry.

For the neutral As vacancy in GaAs ({$V_{As}^0$}), there is just one
electron in the localized state formed by the dangling bonds.  The
volume reduction is more than twice as large as for the positively
charged defect. This change in volume is associated with a Jahn-Teller
distortion with the formation of two dimers, leading to two short
(-11.5\%) and four long Ga-Ga bonds (-24.5\%) arranged in a $D_{2d}$
symmetry. This stretches and weakens the back bonds but it allows all
atoms to recover a fourfold coordination.

Using {\it ab initio} molecular dynamics, Laasonen {\it et
  al.}~\cite{Laa92} see a small (2-3\%) outward relaxation, and an
even smaller (0.6\%) pairing-mode relaxation, leading to a weak
tetragonal distortion with $D_{2d}$ symmetry. This calculation was
found to suffer from band dispersion for the localized defect states
due to the artificial interaction between unit cells; as a consequence
atoms surrounding the vacancy are not allowed to relax properly.
Using tight-binding molecular dynamics, a larger breathing-mode
displacement was obtained by Seong and Lewis.~\cite{Seo95}  They find
that local tetrahedral symmetry was broken, as one neighbor atom of
the defect relaxes inward while the other three relax outward. The
pairing mode relaxation is also found to be very small. Feng {\it et
al.},~\cite{Fen97} using a similar method with a 64-atom supercell,
found similar results leading to trigonal distortion with $C_{3v}$
symmetry.

The difference between our results and previous calculations come from
the use of a better converged potential as well as of a larger unit
cell. Moreover, as stated previously, we have started the relaxation
from various random geometries, always converging to the same final
state: the lowest energy configuration has $D_{2d}$ symmetry for
neutral $V_{As}$.

The extra electron added to get {$V_{As}^{-1}$} can be accommodated in
the same localized level as the previous one.  A stronger pairing mode
relaxation appears and the two short bonds become stronger (from
$-24.5$\% to -31.5\%) atoms of the dimer get closer, while the long bond
are almost kept fixed. In agreement with our results reported in
Table~\ref{tab:relax_As}, Chadi,~\cite{Cha03} using LDA and 32-atom
supercell, finds that the -1 charge state arises from a direct
transition from the +1 state due a pairing of the neighboring Ga atoms
and has $C_{2v}$ symmetry (pairing mode). The Ga atoms would then move
by 0.8 \AA\ ($\sim$ -20\%) from their ideal position to form two sets
of paired bonds.

The next electron added to the localized levels, occupy a different
state. The arrangement of the atoms around the doubly negative vacancy
{$V_{As}^{-2}$} is directly affected: the short bonds remain unchanged
but the long bond become stronger passing from $-12.6$\% to $-17.6$\% with
a slight change in the relaxation volume.  The dimers are therefore
brought closer without affecting the intradimer distance.
 
Most interestingly, the relaxation geometry is modified when a fourth
electron is added. In particular, there is an inversion in the
Jahn-Teller distortion and the pairs of atoms forming the two dimers
get closer to each other and form new weak Ga-Ga bonds with a length
equal to the intradimer distance. Finally a tetramer is formed where
the four Ga atoms are equidistant and fivefold coordinated (three
covalent Ga-As bonds and two weaker Ga-Ga bonds). The tetrahedron
formed by the vacancy's first neighbor has 4 short
bonds and 2 long bond, as can be clearly seen from
Table~\ref{tab:relax_As}.

This type of relaxation, ``resonant bond'' model, was first seen in
calculations for the singly negative divacancy in
silicon,~\cite{Sai94,Lewis96} then for divacancies in
GaAs.~\cite{Poy96} More recently, this relaxation pattern have been
observed experimentally then confirmed using {\it ab-initio} cluster
calculations~\cite{Ogu03} for the $As_{Si}-V_{Si}$ pair in silicon.

\subsubsection{Energetics }
\label{sec:energetics_As}

We calculate the formation energies for all possible charge states of
the As vacancy; the most relevant defects are reported in the first
column of Table~\ref{tab:relax_As}. Most of the earlier calculations
deal with the positively charged vacancy and do not go beyond the $-1$
charge state; there are no recent calculations that report formation
energies for the $-2$ and $-3$ charge states.

In a 32-atom supercell LDA calculation of the formation energy of
charged defects, Northrup and Zhang find that the relaxed
$V_{As}^{+3}$ is stable for both intrinsic and p-type
materials,~\cite{Nor94} with a formation energy 1.7 eV lower than the
(+1) charge state. Our results are in complete disagreement with this
calculation. The formation energy we obtain does not favor the (+3)
charge state under any doping or growing conditions:
$E_{f}^{'}(V_{As}^{+3})$ = 3.5 eV compared to
$E_{f}^{'}(V_{As}^{+1})$= 2.8 eV (in agreement with other
calculations: $E_{f}^{'}(V_{As}^{+1})$= 2.97 eV~\cite{Nor93} using LDA
and 3.09 eV~\cite{Sta99} using SCC-DFTB). This confirms that +1 has
the lowest energy among all other defect.

Figure~\ref{fig:levels_As} displays the formation energies as a function
of Fermi level for various charge states of As vacancies at 0~K.
For p-type GaAs (Fermi level close to the valence band maximum) the +1
charge state has the lowest energy, in agreement  with
recent results from Chadi.~\cite{Cha03}

We note, moreover, 
that lines for the neutral
and the negative charge state intersect before those for the positive
and neutral state; the (0/$-$) ionization level is
located well below the (+/0) level and represents a net signature of level
inversion or of the so-called negative-U effect.~\cite{And75} It is
therefore energetically more favorable to transfer two
electrons at the same time to the defect level from the Fermi level
with the reaction $V_{As}^{+} + 2e \rightarrow V_{As}^{-}$.

Such transfer is associated the strong Jahn-Teller distortion
discussed in Section~\ref{sec:relaxation_As} as the systems goes from
the $+1$ to the $-1$ state. The negatively charged vacancy remains
stable for intrinsic and n-type GaAs but is superseded by the triply
negative vacancy in the heavily n-doped GaAs corresponding to $E_F$ =
1.27 eV (level {\it b} in figure~\ref{fig:levels_As}): the ($-$/$-2$)
and ($-2$/$-3$) defect levels almost collapse, thus the direct
($-$/$-3$) transition is favored. Interestingly, the transition
($-$/$-3$) is associated with a structural change from the pairing
mode relaxation to the resonant mode relaxation as discussed in
Section~\ref{sec:relaxation_As}. The four added electron are paired
two by two as the energy gained from the structural relaxation
overcomes the Coulombic repulsion for each of the two electrons,
supplying a net effective attractive interaction between the
electrons.

\begin{figure*}
\centerline{\includegraphics[width=14cm]{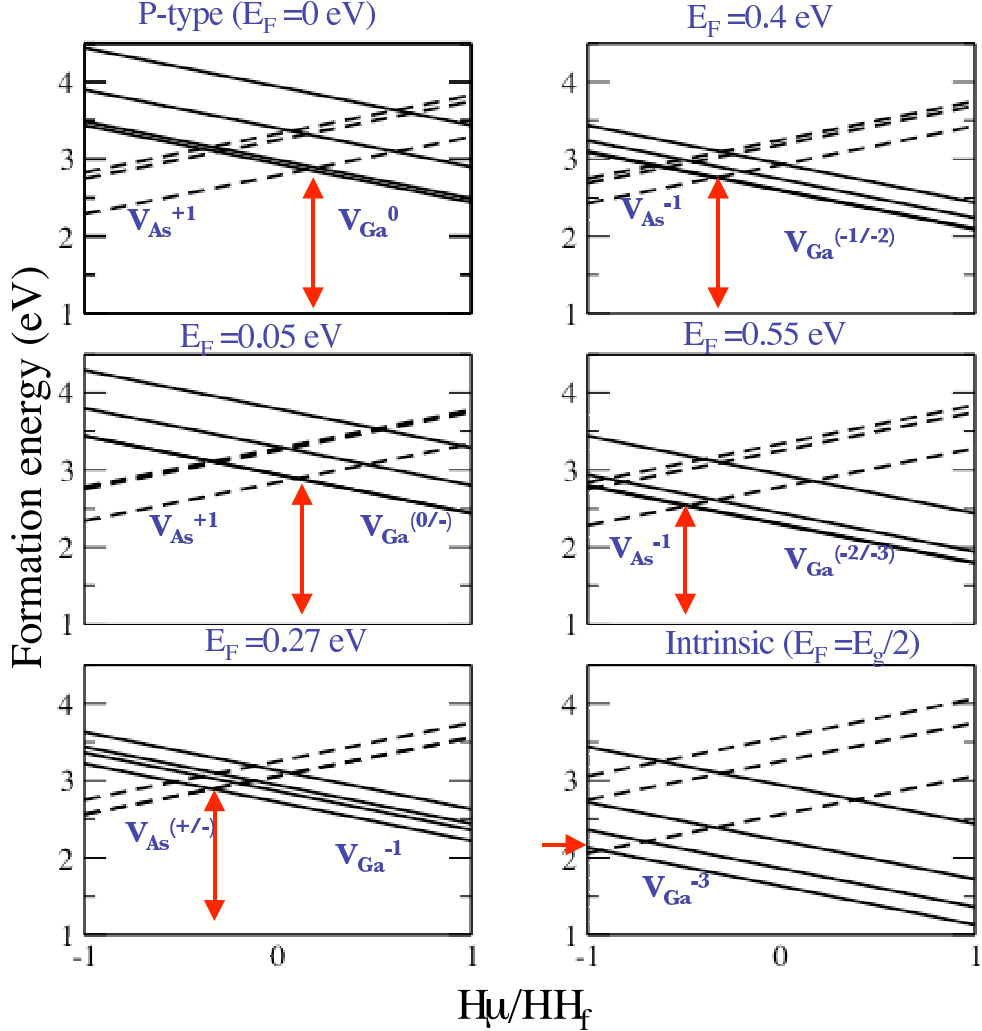}}
\caption{(Color online) Formation energies of Ga (solid line) and
  As (dashed line) vacancies in GaAs as a function of the growth
  conditions ($\Delta \mu$). Different panels are for different
  critical values of the Fermi level or ionization levels identified
  earlier.} 
\label{fig:growth}
\end{figure*}

Only the $+1$, 0 and $-1$
charge states seem to have been studied previously by {\it ab initio}
calculations. 
While Northrup et Zhang~\cite{Nor93}
predict that arsenic vacancy in GaAs exist in the + charge state only,
other calculated ionization levels split into two main categories
classified in Table~\ref{tab:comp_As}: (i) A direct transition from a
charge state to another is possible with only one electron transferred
at time, and (ii) a negative-U effect for the ($+/-$) transition. The
first type of transition is found by Seong and Lewis,~\cite{Seo95}
who predict $V_{As}$ to exist only in the + charge state, at 1.41 eV
above the valence band this charge state changes to neutral then to
negative charge state at the limit of the conduction band maximum. In
an earlier study for unrelaxed As vacancies reported by Jansen and
Sankey~\cite{Jan89} ionization energies are located in the range of
experimental band gap and located near the conduction band maximum. A
negative-U effect is reported by both Cheong and Chang~\cite{Che94}
and P\"oykk\"o {\it et al}~\cite{Poy96} for the ($+/-$) transition, the
transfer of the two electron occurs above the middle of the band gap.
This negative-U behavior is confirmed by a recent calculation from
Chadi~\cite{Cha03} which finds that the direct transition $V_{As}^{+}
+ 2e \rightarrow V_{As}^{-}$ is favored after a Jahn-Teller distortion
and that $-1$ charge state is the most stable for Fermi levels above
midgap.

Our results agree partially with this results. There is a negative-U
effect but the level for the ($+/-$) transition is shallower. This
might be due to the important structural relaxation that affects the
neutral and $-$ charged vacancies. Moreover, as discussed above, we
also find a second negative-U transition level at $E_F$ = 1.27 eV with
the reaction $V_{As}^{-} + 2e \rightarrow V_{As}^{-3}$.

Real GaAs crystals are far from being perfectly stochiometric, during
growth there will be excess of Ga or As ions. A more general study
concerns the effect of the growing conditions on the stability of the
defect under certain doping conditions. In Figure~\ref{fig:growth},
the various panels show the progressive doping of the GaAs sample and
how the stability of the defect gets affected consequently. The Fermi
level ranges between the valence band maximum and the midgap where
most of the ionization levels computed in this work have been
identified. Each of the panels shows the transition between two charge
states for As and Ga vacancies at the critical values of the
ionization levels. For example at $E_f$=0.05 eV the first transition
for gallium vacancies take place, the two line collapse and are
indicated as $V_{Ga}^{(0/-1)}$.

For p-type GaAs, at the As-rich limit in GaAs, corresponding to $\Delta
\mu /\Delta H_f$ close to $-1$, the dominant charge state are the As
vacancies that probably compete with As antisites, while for the
Ga-rich limit Ga vacancies have lower formation energies.  For a Fermi
energy at midgap and for n-type GaAs, regardless of the growing
condition, the triply negative charge state is the most stable among
others and have the lowest formation energy.

\section{Conclusions}

We have presented a complete description of the energetic and the
relaxation geometry for relevant vacancy type defects in GaAs using
the SIESTA {\it ab initio} program.
Various convergence tests show that size effects, the completeness of
the basis sets and the sampling of the Brillouin zone can become very
important when the symmetry of the defect is broken or when the defect
is highly charged. 

Using the DZP basis set, with a 216-atom unit cell and a $2\times
2\times2$ $k$-point sampling, we find that Ga vacancies have shallow
ionization levels below midgap in agreement with experiment, and
do not show any recombination of the dangling bonds as was shown in 
earlier calculation~\cite{Laa92,Seo95}. For the less studied As
vacancy, we find that the ionization level ($+/-$) of As is
located in the lower half of the band gap and lies
 near the valence band while the
second negative-U level ($-/-3$) is located above midgap with a
significant difference in the relaxation pattern reported earlier. The
triply negative charge state for As vacancy reconstruct in the
resonant bond mode, in a similar fashion as divacancies in Si and GaAs
reported earlier.  Finally, we find that only a few vacancy types
can act as vehicles for self-diffusion of dopants in real GaAs devices
under different doping and growing conditions, suvy as the triply negative
Ga vacancy for intrinsic and n-type GaAs. 

These results will be used as a starting point for a detailed study of
self-diffusion using the SIEST-A-RT method presented
elsewhere.~\cite{Elm04b}

\section{Acknowledgments}

We thank Pablo Ordej\'on for his help in the initial optimization of
the basis sets.  This work is funded in part by NSERC, FQRNT and the
Research Canada Chair program. NM is a Cottrell Scholar of the
Research Corporation. Most of the simulations were run on the
computers of the R\'eseau qu\'eb\'ecois de calcul de haute performance
(RQCHP) whose support is gratefully acknowledged.

\bibliography{bibliography_gaAs}
\end{document}